# Influence of buffer/protective layers on the structural and magnetic properties of SmCo films on Silicon

F. Maspero[1], O.V. Koplak[1], A. Plaza[1], B. Heinz[2], F. Kohl[2], P. Pirro[2], R. Bertacco[1]

[1]Dipartimento di Fisica, Politecnico di Milano, Milano, Italy
[2]Fachbereich Physik and Landesforschungszentrum OPTIMAS, Rheinland-Pfälzische Technische Universität Kaiserslautern-Landau, Kaiserslautern, Germany

**Integration of Samarium Cobalt hard magnets on silicon requires buffer/protective layers that can enhance the magnetic properties of the magnet while preserving its structure and chemical composition after post-annealing treatments needed for the formation of the magnetically hard phase. In this work, a comparison of Samarium-Cobalt films for five different buffer/protective layers, namely Ti, W, TiW, Ta, Cr and two different annealing temperatures, 650°C and 750°C, is presented. Depending on materials and annealing temperatures, magnetic properties such as saturation and coercivity of the SmCo film can be finely tuned. We show that coercivity up to 3.65 T or saturation magnetization up to 0.95 T can be reached by proper choice of the relevant process parameters: deposition temperature, material for the buffer/protective layer and annealing temperature. Such value of coercivity is among the highest found in literature for thin films of SmCo.**

## I. Introduction

Rare-earth-based (RE-based) magnetic materials have been extensively studied in the last decades due to their excellent magnetic properties as hard magnets both in the form of bulk materials and thin films. The use of RE magnets at the microscale is of interest for several applications, ranging from recording media [1] to microelectromechanical systems (MEMS) [2]. SmCo is one of the best RE-based compounds in terms of magnetic properties and high Curie temperature. It shows good adhesion on glass or alumina substrate, thanks to the similar expansion coefficient [2] and it was also grown epitaxially on MgO single crystal substrates [3]; however for MEMS application the integration with silicon is fundamental. SmCo usually shows in-plane (IP) crystalline anisotropy [2], [4], [5], [6], [7], [8], [9], [10], [11], nevertheless several works reported on SmCo thin-films with out-of-plane anisotropy [12], [13], [14], [15], [16]. Such properties depend on several parameters, such as deposition or post annealing temperature [4], [9], [13] pressure [9], buffer/protective layers [15], lamination conditions [13], and film thickness. In order to produce samples with stable magnetically hard properties, the conditions for growing SmCo films must be carefully optimized. Table I reports a list of works from literature, highlighting the buffer/protective layers used as well as deposition and annealing temperature. The buffer layers can affect the magnetic properties via several processes and conditions:

- Lattice mismatch between the magnetic and buffer layers. Different buffer layers (Ni, Co, Cu, Pd, Pt, Al, Au) were used for SmCo thin (40 nm) films but only Cu can support the epitaxial growth of $SmCo_5$ (001)-oriented films [17]. In general, the lattice mismatch deeply influences the stress induced on SmCo, even though this becomes less relevant as the thickness increases [18]. Nevertheless it can strongly influence the creation of perpendicular magnetic anisotropy (PMA) [11], [17], [19].
- Mismatch in the thermal expansion coefficient. The high-temperature growth or the post-annealing can induce a large stress in the SmCo film, when cooled at room temperature. This can impact the magnetic properties of the material and its adhesion [10], [20]. For example, a large thermal gradient field in SmCo heterostructures strongly affects the formation of nucleation sites during annealing [21].

TABLE I - TUNING OF MAGNETIC PROPERTIES OF SmCo FILM WITH DIFFERENT BUFFER LAYERS

| Buffer layer | Deposition temperature (°C) | Annealing temp (°C) | Max. coercivity Hc (T) | Reference |
|---|---|---|---|---|
| Cr | RT – 450 | 450 | 5 | * [4] |
| Nb | RT - cooled | 500-750 | 3.48 | [5] |
| Cr | RT - cooled | 560 | 1.5 | [2] |
| TiW/Cu | 300 | 350 | 3.72 | *§ [16] |
| Ru/Cu/Ru | 350 | 350 | 1.32 | *§ [12] |
| Cr | 450 | 550 | 2.7 | [6] |
| Cu | 20–400 | No ann. | 0.76 | *§ [13] |
| Cu-Ti | 325 | No ann. | 1.2 | *§ [14] |
| Cr | RT | 250-750 | 1.55 | [7] |
| Cr - Mo | RT – 450 | No ann. | 1.55 | [8] |
| Ta | RT | 550-650 | 0.8 | [9] |
| Cu-Pt | 350 | No ann. | 1.2 | § [11] |
| Cr | 600 | No ann. | 0.85 | [10] |
| Ni – NiW - W | 530 | No ann. | 1.55 | *§ [15] |

*refers to glass substrate or silicon oxide coated silicon
§ refers to SmCo with out-of-plane anisotropy.
RT stands for room temperature.

- Structural distortion at the interface. This phenomenon can promote the formation of a magnetic dead layer related to the frustrated character of the magnetic configuration [22]. The dead layer is typically 1-2 nm thick and its contribution can be neglected for thick films.
- Interdiffusion at the buffer/SmCo interface. This can contribute to the creation of a magnetic dead layer, as well as promote the orientation of the film grown on top of the interdiffused region. This was observed for SmCo/Cu thin films in [19]. Atoms from the buffer layer can also diffuse in the entire bulk and act as extra element of the compound.
- Thickness of buffer layers. This aspect was investigated

for the SmCo/Cu system in [19]. The values of coercive field and squareness for SmCo/Cu film for out of plane (OOP) fields became larger when increasing the Cu underlayer thickness up to 100 nm, whereas those for in plane (IP) fields showed a drastic decrease.

From this survey it turns out that the search for a suitable material to be used both as buffer and protective layer for SmCo films is crucial for the integration of this material in microsystems.

The focus of this work is to compare the effect on SmCo properties of five different materials (M), namely Ti, Ta, TiW (10/90 wt%), W and Cr, used both as buffer and protective layers in M(100)/SmCo(400)/M(100)/Si structures, where number within brackets indicate the thickness in nm. The choice of TiW, Ta and W is related to their good properties as barrier layer, avoiding interdiffusion and oxidation even after high-temperature processes. Cr was chosen for its well-matched thermal expansion coefficient and as it is the most common buffer found in literature. Titanium was considered as it represents a commonly used adhesion layer for metal deposition and it has been also used as doping material for SmCo [23]. This comparison is carried out with fixed deposition condition and two different post-annealing temperatures, 650 C° and 750 C°, leading to the formation of magnetically hard SmCo films [24]. Finally, to investigate the impact of the specific deposition conditions used in this work, a series of structures with different SmCo thickness were grown using tungsten as buffer/protective layer.

## II. EXPERIMENTAL METHOD

### A. Fabrication

The SmCo films were deposited on 2×2 cm$^2$ coupons of p-type silicon <100> ($\rho$=1-10 $\Omega$·cm) using a sputtering system (Leybold-Heraeus LH Z400 MS). The nominal thickness of the SmCo films is 400 nm, while the thickness of the buffer and capping layers is 100 nm. The films were deposited starting from a base pressure < 5×10$^{-6}$ mbar. Capping and buffer layers were deposited using magnetron sputtering with a power density of 1.1 W/cm$^2$ and argon pressure of 14×10$^{-3}$ mbar. SmCo was deposited by RF sputtering on 3 inches target of nominal composition Sm$_{18}$Co$_{82}$, with a power density of 6.6 W/cm$^2$ and argon pressure 14×10$^{-3}$ mbar. Although depositions were performed at room temperature, self-heating due to plasma raises the temperature during the process, especially during the deposition of SmCo. Each coupon was divided in four pieces of 1×1cm$^2$ and annealed in different conditions to compare the effects of the thermal process on the magnetic properties.

Annealing was performed in vacuum at 10$^{-5}$ mbar pressure using a Rapid Thermal Annealer – UniTemp RTP-150-HV. The temperature was increased using heating ramps of 10 °C/s, while nitrogen was used for cooling; The target temperature was kept for 30 minutes, while cooling took approximately 2 hours. The same procedure was used in our previous work on the optimization of SmCo films with 500 nm thickness using tungsten as buffer layer [24]. Increasing the annealing time up to 60 min did not significantly change the films properties (see Fig. S1 in Supplementary materials).

### B. Sample characterization

The morphology of the samples was analyzed using optical microscopy (Leica INM 200), scanning electron microscopy (SEM LEO 1525 Zeiss) and atomic force microscopy (Keysight 5600LS).

SEM was used to observe the cross-section of the heterostructures, estimate the effective thickness after annealing and to perform composition analysis by exploiting Energy Dispersive X-ray Analysis (EDX).

The crystallinity of the samples was investigated using X-ray diffraction (XRD) (Rigaku SmartLab XE). We used the same measurement protocol for all samples. Speed, step, slits and other parameters were the same for all samples, to allow a direct comparison of spectra.

Finally, the magnetic properties of the SmCo films were measured using a Physical Property Measurement System (PPMS DynaCool) with magnetic field up to 9T.

## III. RESULTS

### A. Morphology

Figure 1 reports one the cross-sections of deposited SmCo films with different buffer/capping layer after annealing at 650 °C and 750 °C. The as deposited film appears amorphous uniform (see Fig.S2 in supplementary materials), while the annealed film develops a granular texture. The film capped with Ta remains compact also after annealing, similar texture is observed for W and TiW. The buffer and capping W layers allow for good adhesion, negligible interdiffusion, and protection against oxidation and chemical corrosion during further film processing as was shown in our previous work [24]. On the other hand, the film capped with Cr and Ti show delamination or formation of cracks and voids in the film as visible in Fig. 1(b), (e), (f). Such degradation is also visible in the AFM analysis reported in Fig. 2.

While for Ta, TiW, and W the whole multilayer maintains a relatively smooth surface, a sizable roughness is found for Ti and Cr. The lowest average roughness parameter ($R_a$) was found for SmCo films with Ta buffer layer. Chromium and Titanium roughness tends to largely increase after annealing. This is also associated to the appearance of cracks and film delamination as mentioned before.

Interestingly enough, our findings are in contrast with some works reporting on the successful usage of Cr and Ti as a buffer layer [2], [4], [6], [7], [10], likely due to the higher annealing temperatures used in our process with respect to that typically used (450°C) in other papers (see Table I). As a matter of fact, diffusion and intermixing at the Cr/SmCo interface after annealing at 500 °C has been previously reported [25]. The higher temperature of our post-annealing could enhance the diffusion of Cr leading to the observed behavior.

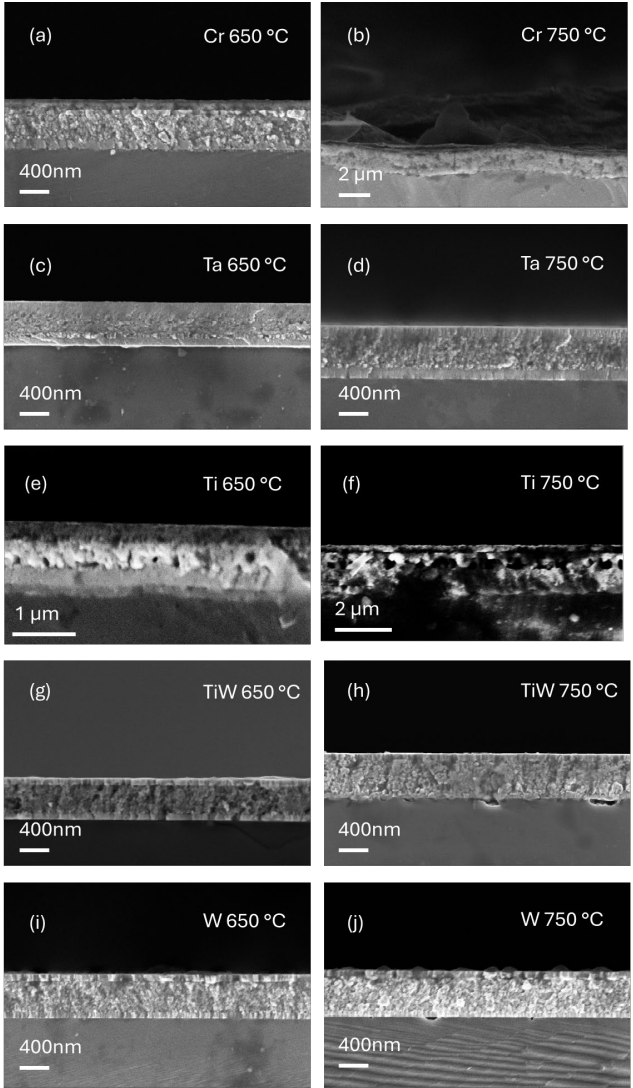

Fig. 1. SEM cross-section image of SmCo at different annealing conditions: Cr 650 °C (a), Cr 750 °C (b), Ta 650 °C (c), Ta 750 °C (d), Ti 650 °C (e), Ti 750 °C (f), TiW 650 °C (g), TiW 750 °C (h), W 650 °C (i), W 750 °C (j).

### B. Composition and crystal structure

The composition of the films was analyzed by means of EDX, confirming a ratio between Sm and Co close to 4.8 for all SmCo films grown on different buffers, in line with the target composition. However, we should keep in mind that EDX estimates cannot be used for fine quantitative analysis of composition as they also depend on the film morphology, which in our case is related to the buffer. Therefore, we decided to exclude a comparison of different buffer in terms of EDX composition.

To investigate the crystal structure, we measured XRD θ−2θ scans on all heterostructures: Cr/SmCo/Cr, Ta/SmCo/Ta, Ti/SmCo/Ti, TiW/SmCo/TiW and W/SmCo/W (Fig. 3(a)-(e)). All samples display features which can be attributed to two magnetic hexagonal phases: $SmCo_5$ (P6/mmm, lattice parameters $a = b = 8.25$ Å, $c = 8$ Å) and $Sm_2Co_{17}$ (P 63/mmc, lattice parameters $a = b = 4.96$ Å, $c = 3.96$ Å). The nominal positions of the main peaks from these two phases, together with those arising from each adhesion layer, are shown in Fig. 3.

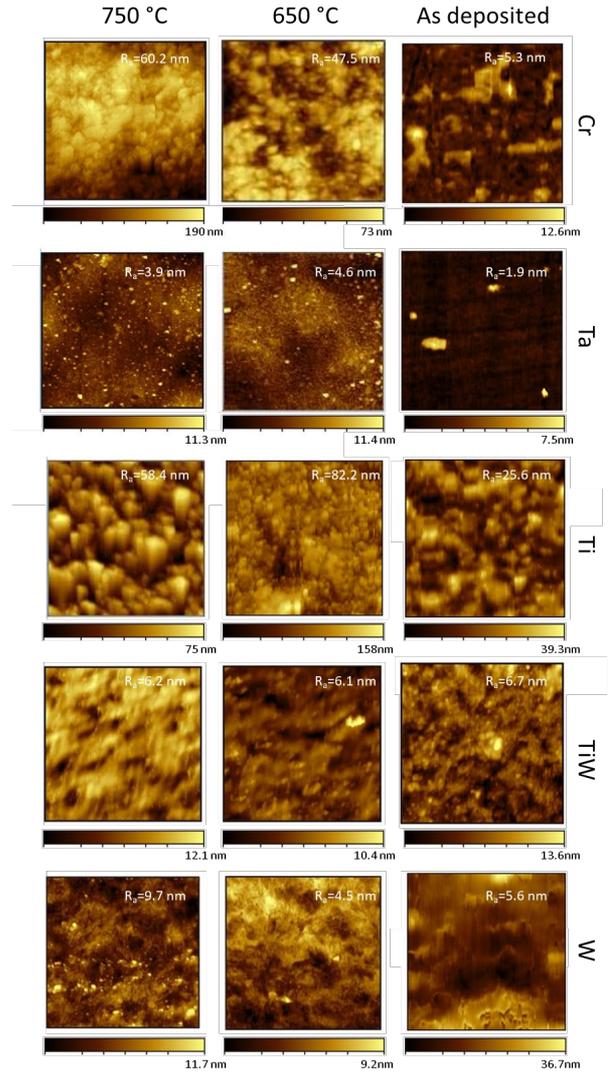

Fig. 2. AFM images of the surface of the film for each buffer/capping layer before and after annealing. All AFM images size of 3 × 3 μm².

In panel Fig. 3(a) θ−2θ scans from Chromium buffered samples are shown. The first visible result is the disappearance of the chromium (110) peak after annealing, which clearly indicates interdiffusion and loss of the polycrystalline structure of the as-grown Cr. When increasing the post-annealing temperature from 650°C to 750°C we observe a redistribution of the spectral weight, showing an overall enhancement of $SmCo_5$ features and a change in the texture of the $Sm_2Co_{17}$ phase, as indicated by the sizable enhancement of the $Sm_2Co_{17}$(104) and $Sm_2Co_{17}$(213) reflections at ~ 48°. This can be ascribed to the diffusion of Cr in SmCo where, it is supposed to partially substitute Co. The atomic radius of Cr (0.166 nm) is larger than that of Co (0.152 nm), so that a distortion and reorientation of SmCo atomic planes can be easily understood.

Figure 3(b) shows the XRD spectra from heterostructures with Tantalum as buffer layer. Already in as-grown structures Ta develops a mixture of the tetragonal β-phase (space group P42/mnm, lattice parameters $a = 1.0194$ nm, $c = 0.5313$ nm.) and bcc phase (I m -3 m space group, lattice parameter $a = 3.3$ Å). The presence of a prominent Ta-β(200) peak at ~33.5° indicates a predominant β-phase with the (100) planes preferentially oriented parallel to the substrate [26]. Pole Figure (PF) measurements were made by recording the intensity of a Tantalum Bragg reflection β-Ta(200) and β- Ta(410) as a function of rotation and tilt of the sample. PF measurements have been carried out for the observation of preferential crystalline orientation of Tantalum (see Supplementary materials Fig. S3). The diffracted intensity was collected by varying two geometrical parameters, i.e. the α angle (tilt angle between the scattering plane and the sample normal) from 0 to 90° and the β angle (azimuth) from 0 to 360°. The diffraction angle 2θ was fixed at 33.2° corresponding to the maximum of the Ta-β (200) (Fig. S3(a)) and at 36° corresponding to the maximum of the and Ta-β (410) reflection (Fig. S3(b)). Pole Figure measurements for β-Ta(200) and β- Ta(410) peaks confirmed texturization (see Supplementary materials Fig. S3). Upon annealing at 650°C we observe a small shift of the β-Ta(200) peak by 0.5° towards smaller angles, indicating a change in the cell parameter, together with the appearance of characteristic features of polycrystalline SmCo at ~ 43°. The observed change of the Ta lattice parameter upon annealing can be explained in terms of the stress induced by the crystallization of the SmCo film. When increasing the annealing temperature at 750°C we observe instead a strong reduction of the β-Ta(200) peak accompanied by some increase of the β-Ta(410) peak. This could indicate a re-orientation of Ta grains as well some interdiffusion promoted by the high-temperature annealing. In parallel we observe a sizable change of features related to SmCo, with the appearance of additional peaks (like that at 45° which can be ascribed to $Sm_2Co_{17}$(302) reflection) indicating a major reorientation of grains.

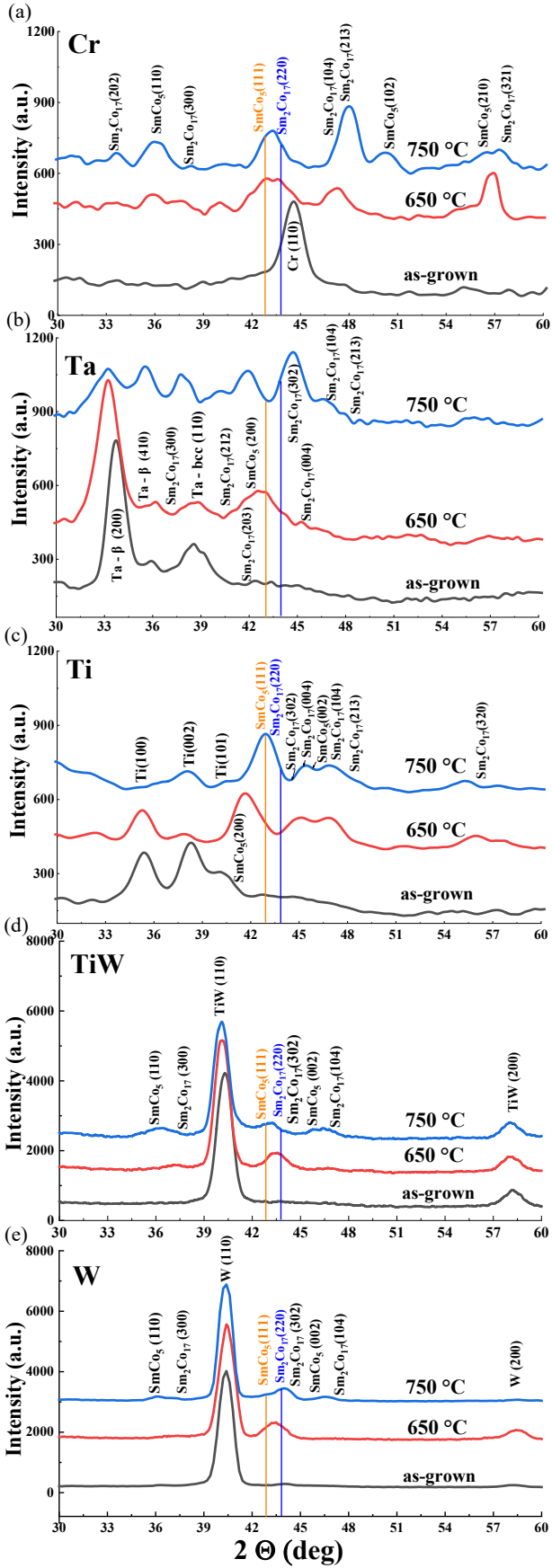

Fig. 3. XRD spectra of the SmCo films with different buffer layers: Cr (a), Ta (b), Ti (c), TiW (d) and W (e)

The XRD analysis of Ti/SmCo/Ti structures is reported in Fig. 3(c). As deposited films display features coherent with the characteristic hexagonal close-packed structure (hcp) of Titanium. The presence of (100), (002) and (101) reflections indicates that the film is polycrystalline with a weak texturization. After annealing at 650°C, as for Chromium, the Ti(002) peak almost disappears and also the Ti(100) is strongly attenuated. This reflects the degradation of the stack observed by SEM (see Fig. 1(c)-(d)). At 750°C this phenomenon is enhanced, and the Ti(100) almost disappears. In parallel, we observe the appearance of characteristic features from $SmCo_5$ and $Sm_2Co_{17}$. Noteworthy the buffer layer stimulates a peculiar orientation of grains, with a sizable peak at 42° (corresponding to $SmCo_5$(200) grains) visible at 650°C which transforms into another peak at 43° at 750°C. This second peak, which can be mainly ascribed to $SmCo_5$(111) reflection, together with other features between 45° and 48°, correspond to the fingerprint of the polycrystalline mixture of $SmCo_5$ and $Sm_2Co_{17}$ phases obtained at high temperature for all buffer layers, with the unique exception of Ta. However, the degradation of the stack with Ti as buffer/capping layer during annealing makes this material unsuited for SmCo integration. Titanium showed poor adhesion properties and Ti/SmCo/Ti films peeled off.

The analysis of TiW and W-coated films shows a clearer situation. The peaks related to the buffer layers are well preserved even after annealing at 750°C, clearly pointing out the chemical and structural robustness of these stacks. Indeed, the XRD spectrum of TiW/SmCo/TiW film shows a prominent (110) peak and a minor one (200) which indicate a sizable film texture (see the pole figure analysis in the supplementary materials Fig. S4). For TiW the (110) is just slightly shifted to smaller angles upon annealing, thus suggesting that the crystallization of SmCo is inducing some additional stress, as for the Ta layer (Fig. 3(b)). After annealing, the amorphous "as-grown" SmCo film transforms to a mix of polycrystalline phases in both films (Fig. 3(d)-(e)), with a XRD spectrum like that found for films grown on W [24]. The spectral weight slightly changes depending on the buffer material and annealing temperature. Overall, the SmCo film appears as a mixture of $SmCo_5$ and $Sm_2Co_{17}$ phases, with an increasing weight of the first one upon annealing at 750°C which is connected to the magnetic hardening of the film.

### C. Magnetic properties

The magnetic properties of different stacks have been investigated by VSM in the ±9 T range of applied fields. The following analysis does not include the films with Ti adhesion layers and with Cr adhesion layer annealed at 750° due to their poor magnetic signal and/or problems of delamination making them unsuitable for integration in microsystems. Furthermore, the loops are presented as measured without correction of the diamagnetic background of the Si substrate. All films showed a soft ferromagnetic behavior before annealing. This is evident from Fig. 4 showing a typical in-plane and out-of-plane hysteresis loop of the as-deposited W/SmCo/W/Si heterostructure.

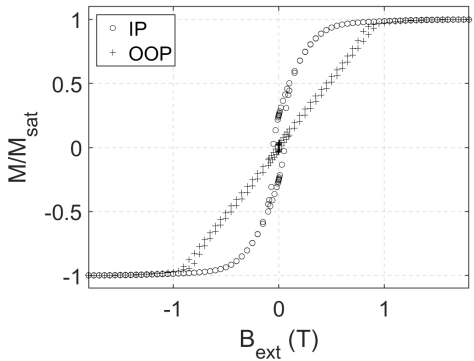

Fig. 4. Typical IP and OOP hysteresis curves of as-deposited W/SmCo/W film. Similar behavior is observed in all films. This analysis was performed between ±2 T.

The hard magnetic properties of SmCo appear after annealing, as shown in Fig. 5 and 6, always for the W/SmCo/W/Si stack. The full set of hysteresis loops for all samples can be found in the Supplementary materials Fig. S5. The values of magnetization are given in tesla, by calculating the magnetic volume as follows: the effective surface was computed using optical images of the sample, while the thickness was assumed constant and equal to 400 nm from SEM cross-sections. This method requires to discriminate the interface between the buffer and the SmCo, which sometime is not trivial, especially after annealing. Moreover, the non-uniformity of the deposition tool can contribute to error in the estimate of the thickness, which however should remain within ±10%. The absolute value of the magnetization reported below are therefore subject to the same uncertainty, yet the trend between remanence and coercivity is still valid, especially for samples with the same buffer layer/coming from the same coupon.

Starting with the in-plane configuration (Fig. 5), the coercive field reaches a value of 0.38 T, upon annealing at 650°C (orange curve) and up to 3.65 T upon annealing at 750°C (blue curve). In parallel, the saturation magnetization decreases from 0.94 T to about 0.5 T. This is consistent with the general trend observed in SmCo films, whose coercivity (magnetization) increases (decreases) with the ratio Sm/Co. The large coercive field is due to the development of $Sm_2Co_{17}$ and $SmCo_5$ phases and we know from XRD that annealing at 750°C promotes the formation of the Sm-rich $SmCo_5$ phase. The loops display a small step close to zero field, corresponding to a soft amorphous phase, while at least two other "steps" are present, at higher fields. The first can be associated to the $Sm_2Co_{17}$ phase, while the second can be attributed to the switching of $SmCo_5$ grains exchange coupled to the softer phase ($Sm_2Co_{17}$) [27]. To further show the presence of steps we added a graph of the first derivative of the magnetization with respect to the external field (Fig. 5(b)-(c)).

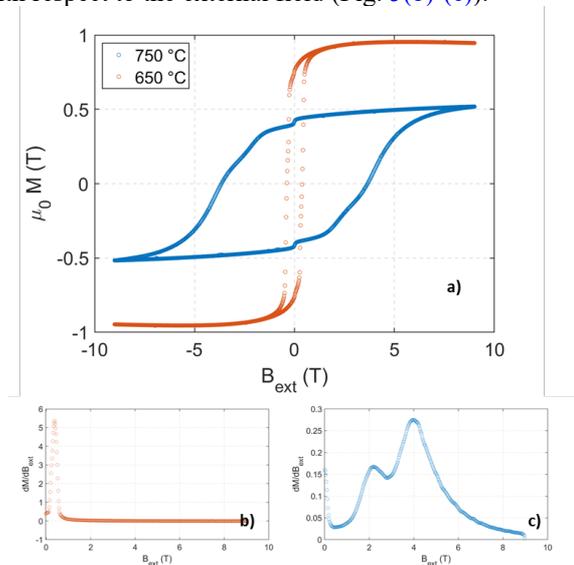

Fig. 5. (a) IP hysteresis curves for W/SmCo/W after annealing at 650° and 750°. Derivative of $dM/B_{ext}$ for (b) the sample annealed at 650 °C and (c) the sample annealed at 750 °C.

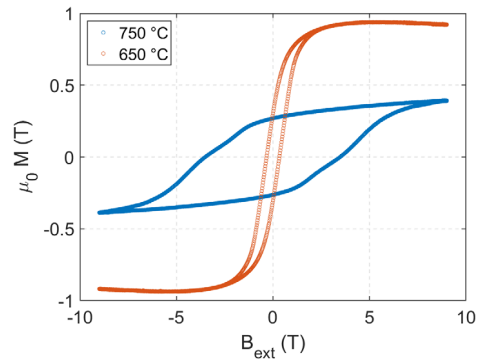

Fig. 6. OOP hysteresis curves for W/SmCo/W after annealing at 650° and 750°.

A similar situation is found for OOP loops (Fig. 6), even though here the loops are less squared with respect to IP ones (Table 2). This is likely due to a predominant in-plane orientation of the easy axis of the magnetocrystalline anisotropy and partially due to the shape anisotropy of the film. In our case, some OOP loops do not show a full saturation of the film, which means that the effective anisotropy field must be larger than 9 T. This is not consistent with a pure shape anisotropy effect as in this case the anisotropy field cannot exceed the saturation magnetization, which is on the order of 1 T.

TABLE II - SQUARENESS RATIO OF EACH SAMPLES FOR IP AND OOP LOOPS

| $M_r/M_s$ | Ta 650 | Ta 750 | TiW 650 | TiW 750 | W 650 | W 750 | Cr 650 |
|---|---|---|---|---|---|---|---|
| IP | 0.85 | 0.82 | 0.79 | 0.80 | 0.80 | 0.81 | 0.73 |
| OOP | 0.58 | 0.52 | 0.49 | 0.65 | 0.32 | 0.68 | 0.67 |

On the other hand, it is well known that SmCo magnets feature a large magnetocrystalline anisotropy field, which can be as high as $H_A = 53$ T, an order of magnitude higher than pure Co [28]. In another work, for $SmCo_5$ micromagnets the easy axis was found parallel to the c-axis and the anisotropy field of was found to be ~ 40 T [29]. Since these values are much larger than the 9 T applied in the OOP configuration during our measurements, we are indeed expected not to see a full film saturation if there is a prevalence of grains with in-plane easy axis.

To compare the impact of the different buffer/protective layers, we filled a scatter plot showing for each layer and temperature the magnetic performance expressed in terms of saturation magnetization and coercivity, both for in-plane and out-of-plane magnetization (Fig. 7 and Fig. 8 respectively).

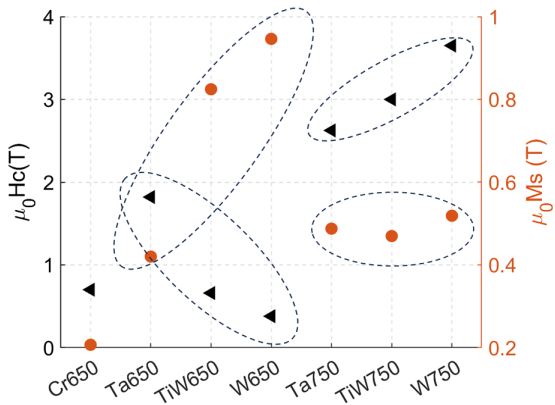

Fig. 7. Comparison of $H_c$ and $M_s$ for different buffer layers and annealing temperatures for IP direction.

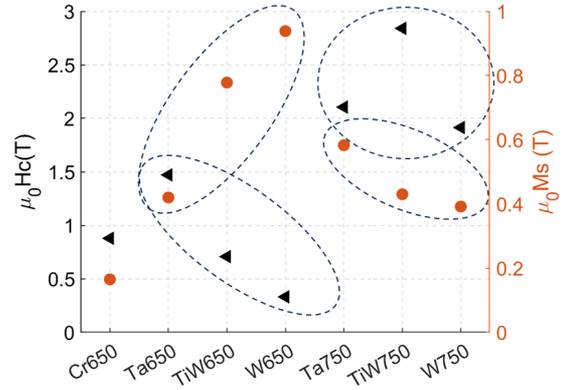

Fig. 8. Comparison of $H_c$ and $M_s$ for different buffer layers and annealing temperatures in the OOP direction.

All films showed good magnetic performance, except for those using Chromium as buffer layer, displaying poor saturation and coercivity already at 650°C. This result is not in line with literature, where various works indicate chromium as optimal buffer layer. However, as seen in Table 1, the temperature used both for the deposition and annealing (450°C) is much lower than the minimum (650°C) required for the recrystallization of films grown at room temperature in the present work. Our results confirm the existence of a threshold temperature (lower than 650°C) above which the intermixing between SmCo and Cr is highly detrimental.

Regarding the other buffer layers, a trade-off between saturation and coercivity is present in all samples, in line with literature [7], [8], especially with the work of Sapktota et al., who showed a clear change in the shape of the hysteresis loop as a function of annealing temperature. Such behavior can be explained with the gradual transformation of $Sm_2Co_{17}$ into $SmCo_5$. This clearly happens when increasing the annealing temperature, as suggested by the XRD spectra of Fig. 3, and can explain the systematic increase (decrease) of the coercive field (saturation magnetization) reported in Figs. 7 - 8 for Ta, TiW and W buffer layers from 650°C to 750° [8]. On the other hand, the observed trend at fixed annealing temperature when moving from Ta to W needs a more subtle explanation.

First, note that films annealed at 650 °C show a larger change in terms of saturation-coercivity with respect to those annealed at 750 °C. This is because 650 °C is the temperature where the transition between the soft amorphous phase and the hard phase starts to appear. Therefore, any slight change in the system can induce drastic change in the magnetic properties. Films annealed at 750 °C instead display a poorly variable saturation magnetization while the coercivity is quite different, with multilayers involving W (TiW) showing the largest value in the IP (OOP) configuration. The trends in the magnetic performance of films annealed at same temperatures are highlighted in the graphs with dashed ellipses.

To explain such a behavior, we must consider that the magnetic properties of SmCo depend on several parameters, such as temperature reached by the film during the growth, type of buffer layer and post-annealing temperatures. While we carefully control the annealing temperature, the same does not apply to the growth temperature as the films are grown without heating but without an active temperature control. In some calibration experiments we estimated that the surface temperature can reach up to 300 °C in conditions of sputtering like those used in the present work, depending on the thermal transmission coefficients of the substrate and sample holder as well as on the deposition time.

To try to disentangle the effect of buffer layers from the deposition conditions we have deposited a set of films using W as buffer layer, while changing only the thickness (Fig. 9). Figure 9 shows the coercivity versus thickness of different W/SmCo(t)/W films, where the thickness $t$ ranges from 50 to 1000 nm. In the very same deposition conditions, thin films show soft-magnetic behavior, pointing to the fact that the very first layer of SmCo are grown "cold" and do not show hard magnetic properties.

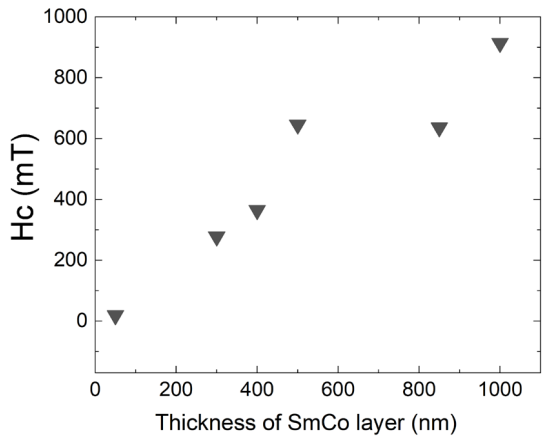

Fig. 9. Coercivity versus thickness of W/SmCo/W films annealed at 650°C. Red line is the linear fit to the data. The trend suggests that the self-heating of the sample during the growth induces the formation of different SmCo phases and coercivity mechanisms.

The main reason for this behavior is not an interfacial dead layer, as we observe a transition of the whole film when increasing the thickness. Instead, we suggest that the observed behavior is related to the increase of the film temperature for larger thicknesses, which are associated to longer deposition times. As the growth continues the film is heated up by the plasma [30] and a partial crystallization is expected to take place already in the as-grown film. By consequence, SmCo phases showing larger coercivity appear in thicker films upon annealing, reaching a coercivity up to 1T. This experiment shows that the final properties of the materials are a combination of the temperature during deposition and the post annealing temperature. This suggests that a possible explanation of the trend observed in films of equal thickness annealed at the same temperature could be caused by the thermal conductivity of the buffer layer, which defines how much the film heats up during growth. The trend observed in the coercivity is the same as observed in the thermal conductivity of the buffer. At room temperature, tungsten has a thermal conductivity around 170 W/(m·K) while the one of tantalum is below 60 W/(m·K) [31], [32]. TiW is expected to have a thermal conductivity in between tantalum and tungsten as it is mainly tungsten, but with a 10% wt of titanium which has low thermal conductivity. During deposition the energy received by the plasma is dissipated differently from the sample surface to the substrate depending on the buffer/protective layer. For Ta, which displays the lowest thermal conductivity, we expect the highest surface temperature during SmCo growth, thus reflecting in a larger thermal budget that promotes the formation of the $SmCo_5$ hard phase upon annealing. This phenomenon can combine also with the different thermal expansion coefficients of tantalum and tungsten, which are respectively 6.6 m/(mK) and 4.6 m/(mK) at RT respectively, thus leading to a different stress condition of the films [33], [34].

IV. CONCLUSIONS

Different buffer/protective layers for the integration of SmCo on silicon have been compared in terms of morphological, crystalline, and magnetic properties. Titanium and chromium suffer from interdiffusion and failure in film passivation which in turn lead to loss of magnetic properties and delamination. Tungsten, titanium-tungsten and tantalum show good performance from a morphological and magnetic point of view. The differences observed in the magnetic properties between these three buffers are larger for films annealed at 650°C, where the development of the hard phase can be strongly influenced by the deposition conditions. In this sense, the possible impact of different buffers can be linked to different thermal conductivity which impact the formation of crystal grains during growth and post annealing processes. On the other hand, films annealed at 750°C show a more uniform behavior in terms of saturation magnetization. Overall, magnetic properties are defined by an interplay between temperature of the film during growth, type of buffer and post annealing temperature. We can conclude that for films grown at low temperature and crystallized by post- annealing to obtain the hard phase, the best buffer/protective layers among those investigated in the present work are Ta, TiW and W. Depending on the specific applications, different combinations of buffer layers and annealing temperatures can be used to finely tune coercivity and remanence in a wide range of parameters.

ACKNOWLEDGMENTS

This work was carried out in the frame of the European Project M&MEMS. The M&MEMS project is funded by the EU under the Horizon Europe programme (contract number: 101070536). This work was also supported by the Joint Research Centre MEMS between Politecnico di Milano and STMicroelectronics. The authors acknowledge the availability of experimental facilities at PoliFAB, the Micro- and Nanotechnology Center of the Politecnico di Milano.

OPEN DATA

Open data are available on Zenodo at the following link: https://zenodo.org/uploads/11160304

# Supplementary materials

**Influence of buffer/protective layers on the structural and magnetic properties of SmCo films on Silicon**

F. Maspero[1], O.V. Koplak[1], A. Plaza[1], B. Heinz[2], F. Kohl[2], P. Pirro[2], R. Bertacco[1]

[1]*Dipartimento di Fisica, Politecnico di Milano, Milano, Italy*
[2]*Fachbereich Physik and Landesforschungszentrum OPTIMAS, Rheinland-Pfälzische Technische Universität Kaiserslautern-Landau, Kaiserslautern, Germany*

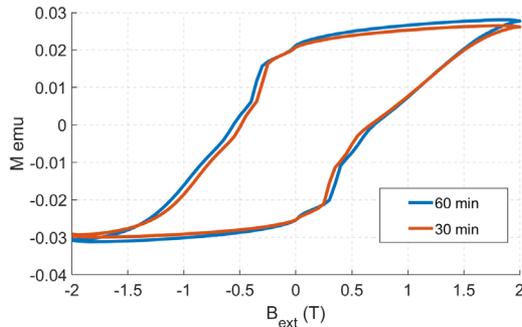

*Figure S1 – IP hysteresis curves for W/SmCo/W after annealing at 650° for 30 and 60 min.*

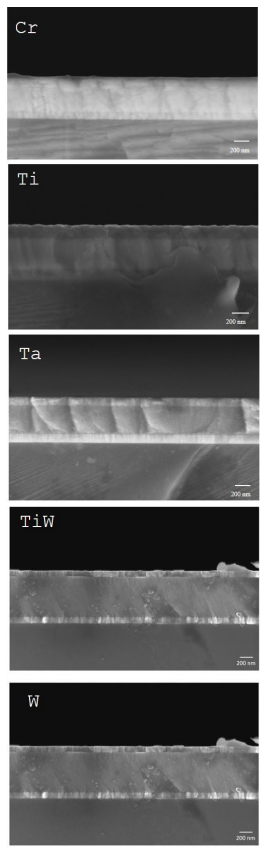

*Figure S2 – cross section of all SmCo films with different buffer layers before annealing.*

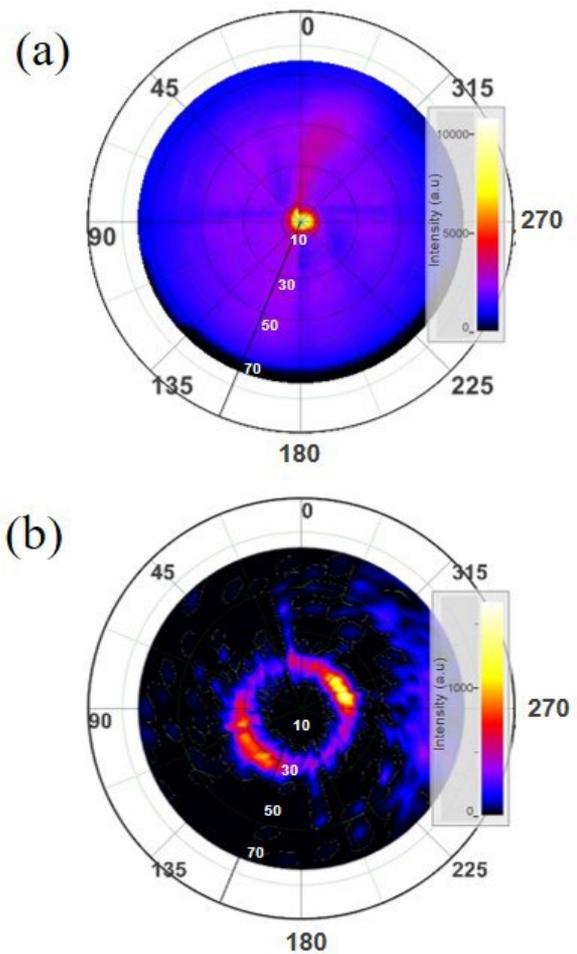

*Figure S3 – Pole Figure for the Ta-β (200) (a) and Ta-β (410) (b) after annealing at 650 °C. The diffraction angle 2θ was fixed at 33.2° corresponding to the maximum of the Ta-β (200) and at 36° corresponding to the maximum of the and Ta-β (410) reflection.*

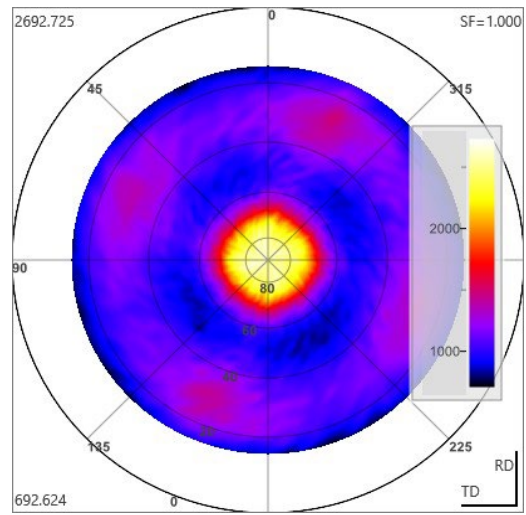

*Figure S4 – Pole Figure for the W(110) after annealing at 650 °C. 650C and 750 C for 30 min. The diffraction angle 2θ was fixed at 40.4° corresponding to the maximum of the W(110)*

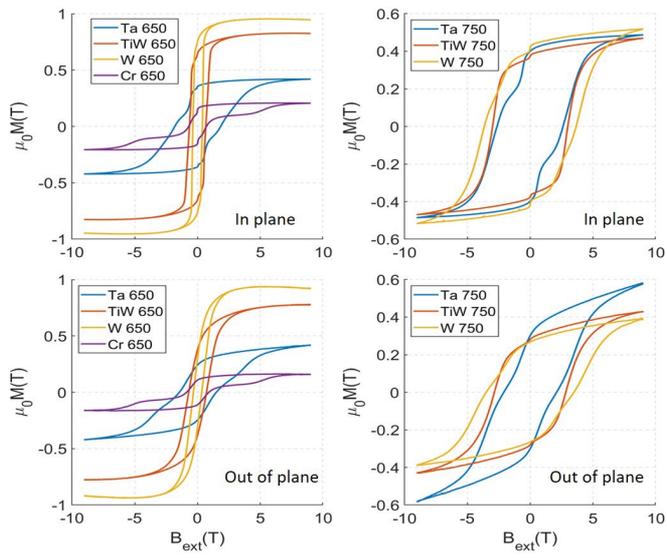

*Figure S5 – IP hysteresis curves for W/Ta/TiW-capped films annealed at 650° (left) and 750°C (right) and Cr-capped film annealed at 650 °C. (top); OOP hysteresis curves of the same samples (bottom).*